# Motion-Based Self-Normalisation Enables Precision Nanoparticle Sizing

D-Dré K.J.M.J. Braam[1,2], Daan Wolters[1], Matz Liebel[1,*]

[1] *Department of Physics and Astronomy, Vrije Universiteit Amsterdam, De Boelelaan 1100, Amsterdam, 1081 HZ, The Netherlands*

[2] *Present address: Faculty of Medical Sciences, Radboud University, Geert Grooteplein Noord 21, Nijmegen, 6525 EZ, The Netherlands*

** email: m.liebel@vu.nl*

**ABSTRACT**

Label-free all-optical nanosizing based on interferometric or darkfield-imaging infers particle size, composition, or shape from single-particle scattering signals. These signals are, however, scaled by a spatially varying normalisation profile that combines the non-uniform illumination with position-dependent detection and data-processing effects. Common normalisation strategies estimate this profile from a direct measurement of the illumination field or background. This is not possible in background-free geometries, such as side- or light-sheet-type illumination, where the illumination does not reach the detector. Here we introduce a self-normalisation method that reconstructs the normalisation profile from the scattering signals of freely diffusing nanoparticles, without additional hardware or calibration samples. To remove the bias caused by particle-to-particle heterogeneity, single-particle trajectories are normalised against each other in regions of spatial overlap, where distinct particles sample the same normalisation profile. In a transmission geometry, the reconstructed profile agrees qualitatively with a direct illumination measurement and reduces signal variations within single trajectories from 6.5% to 2.0% (one sigma), corresponding to a diameter uncertainty of about 0.7%. Under 90° side-illumination, where the illumination field cannot be measured through the detection path, normalisation with the reconstructed 3D profile reduces within-trajectory signal variations from about 30% to 4–10% across most of the sampled volume. As the method solely relies on the scattering signal already acquired, our approach is directly compatible with existing interferometric and darkfield nanoscopy platforms and extendable to other scattering or fluorescent modalities, including light-sheet microscopy.

## INTRODUCTION

Methods for the label-free sizing and characterisation of nanomaterials are highly relevant across many disciplines. High-tech solutions such as cryo electron microscopy or analytical ultracentrifugation provide impressive results but come at a considerable price tag combined with low-throughput[1]. Optical alternatives such as dynamic light scattering or nanoparticle (NP) tracking analysis are much cheaper and compatible with high-throughput screening but compromise on characterisation accuracy and precision[2,3]. All-optical interferometric, or holographic, single particle microscopes, such as interferometric scattering microscopy, sometimes referred to as mass photometry,[4] or off-axis holography[5–7] promise better solutions by retaining the cost and throughput advantage while improving accuracy and precision. Several solutions have already entered the nanosizing market targeting surface based ultrasensitive single-protein characterisation (*Refeyn TwoMP*) and free-flow nanocharacterisation (*Bruker iNTApharma*). In the context of nanosizing, a crucial metric is sizing precision. A highly precise instrument could be used to *in-situ* study antigen-antibody interactions or distinguish accurately loaded viral delivery systems from those with partially damaged nucleic acid load. Achieving this ambitious goal requires high-quality signal normalisation and correction schemes, an area of intense ongoing research. Illumination geometries, experimental noise, optical transfer functions and aberrations are some of the parameters that impact the acquired signals[8–10]. An often overlooked parameter is the impact of the spatially non-uniform illumination intensity which is often estimated through median backgrounds or differential imaging schemes. These provide good approximations as long as straylight that does not originate from the sample plane can be excluded.

Figure 1a schematically outlines the normalisation challenge that we will be addressing in this manuscript by depicting a simplified, minimal-complexity, experiment. Particles of interest are illuminated with an illumination field, $E_{illu}$. Upon light-matter interaction, the particles scatter light, $E_s$, which is then collected and imaged onto a camera. For holographic technologies a reference field, $E_r$, interferes with $E_s$ at the camera resulting in an image, $I_{holo}$, that can be described as follows:

$$I_{holo} = A_s^2 + A_r^2 + A_s A_r \left(exp^{[i\Delta\varphi]} + exp^{[-i\Delta\varphi]}\right) = A_s^2 + A_r^2 + 2A_s A_r \cos[\Delta\varphi] \qquad (1)$$

With $A_s$ being the amplitude of the scattered electric field, $A_r$ that of the reference field and $\Delta\varphi$ the phase difference between the two fields. A key challenge is that $E_{illu}$ exhibits a spatially non-uniform amplitude, $A_{illu}$. Furthermore, position dependent optical transfer functions (OTFs)[11] might impact how $E_s$ is mapped onto the detector plane which can directly impact $A_s$ but also the quantity inferred through data-processing which might be sensitive to optical aberrations. In simple terms, even if all the particles in Figure 1a were identical, the signals recorded for the individual particles would differ dramatically. The challenge is thus to recover the correct signals by normalising for the spatially dependent $A_{illu}$ and the impact of the OTF on the signals as extracted through the data-processing algorithm of choice. Throughout this paper we refer to the combination of these spatially dependent factors as the normalisation profile.

For some approaches, such as inline holographies, it is possible to estimate $A_{illu}$ by, for example, acquiring a median background and assuming that parasitic back-reflections are absent. OTFs are often assumed to be spatially independent, an assumption that can be justified for small fields of view. Depending on the level of sizing precision required, both assumptions can become increasingly difficult to meet. Other schemes relying on off-axis holographic or darkfield imaging bring additional challenges as they allow arbitrary illumination geometries. Figure 1b shows the example of a background-free side-illumination configuration that we favour in off-axis implementations. In such a case, even coarsely estimating the normalisation profile is non-trivial as the illumination light is not entering the detection path, the main reason why this approach yields excellent background suppression.

Ultimately, sizing accuracy and precision are directly correlated with the normalisation quality (Figure 1c). An approach that intrinsically accounts for the normalisation profile and is robust with respect to parasitic back-reflections while being widely applicable to all illumination geometries would be highly desirable. In this manuscript, we propose an easy to implement self-normalisation scheme that extracts the normalisation profile via the same quantity that is of central interest in nanosizing: light scattering by single NPs which makes our approach highly compatible with all existing implementations of interferometric and darkfield-based nanoscopy.

The manuscript is organised as follows. First, we provide a thorough discussion of the normalisation problem and describe our approach using an intuitive 2D sample configuration that compares uniform and non-uniform particle populations. Next, we demonstrate that eliminating intrinsic particle size heterogeneity is essential for successful normalisation, and explain how to achieve this despite the inherent heterogeneity of essentially all nanomaterials. We then extend our framework to freely moving particles in 3D volumes. Finally, we provide several validation experiments to showcase our approach, explicitly highlighting its performance in cases where it is impossible to directly measure the illumination field. Throughout the manuscript, we report holographically measured scattering signals acquired using a "darkfield" type detection, analogous to Figure 1a,b where the collected scattering signals interfere with an external reference wave. A setup-schematic, analogous to[12,13] is presented in the Supplementary Information 4.

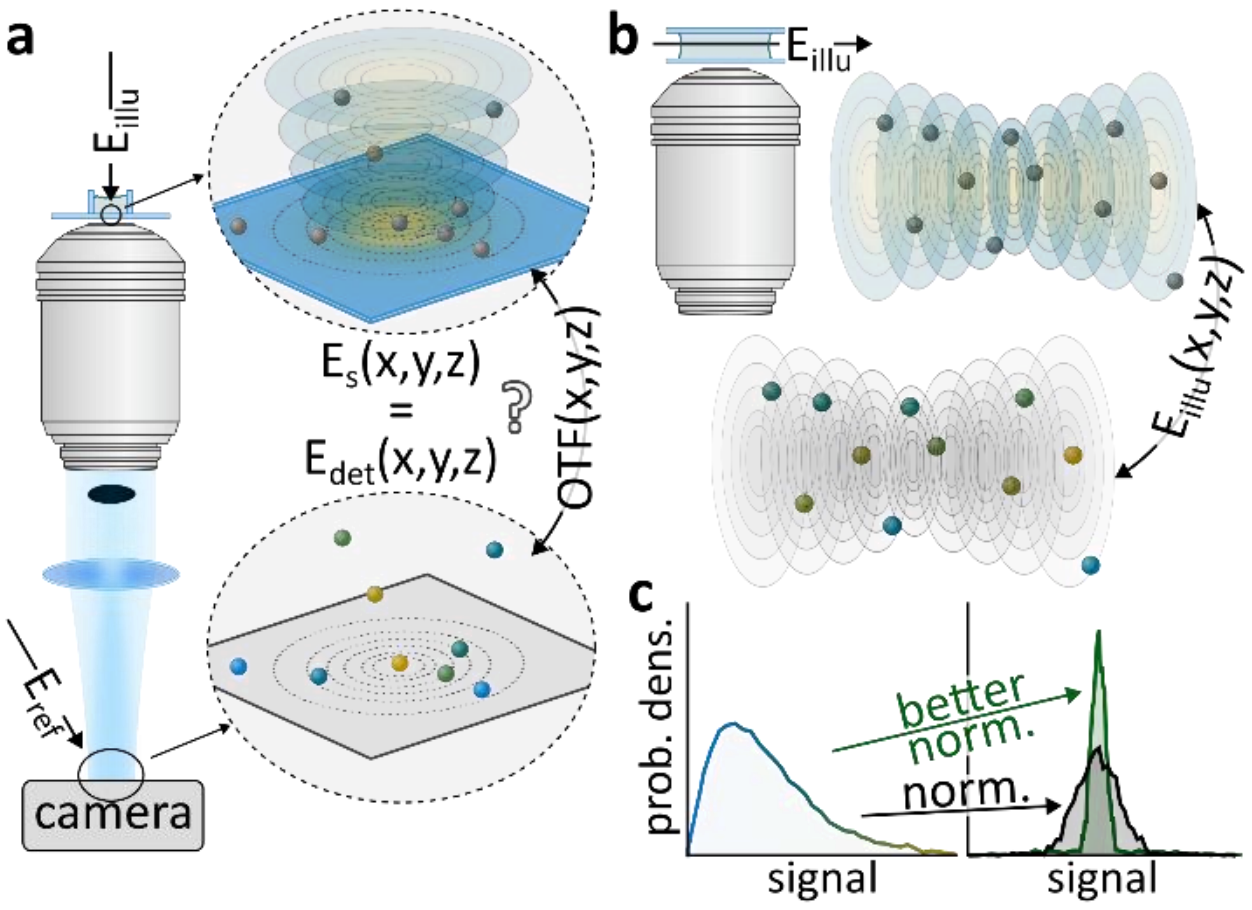

**Figure 1, The signal normalisation challenge in interferometric nanoscopy.** a) A simplified holographic sizing experiment assuming identical NPs. Even though the sample is monodisperse, the collected scattering signals, $E_s$, exhibit significant variability. This is due to two factors: the spatially non-uniform illumination, $E_{illu}$, and the system's spatially dependent optical transfer function (OTF). b) Example of an illumination geometry that makes retrieving $E_{illu}$ highly non-trivial. c) A fictional measured signal-distribution for nominally identical particles (blue) requires high-quality normalisation to yield signal-distributions resembling ground truth (black, green).

## RESULTS

The challenge of any scattering-based nanosizing technology is to extract $A_s$ from the measurement and convert it into a quantity that directly reports on the size or mass of the analytes of interest.

Consulting the general signal (Equation 1) we first need to eliminate $A_r^2$ which can be achieved by, for example, acquiring a median background:

$$I_{holoclean} = A_s^2 + 2A_sA_r\cos[\Delta\varphi] = A_s^2 + A_sA_r\left(exp^{[i\Delta\varphi]} + exp^{[-i\Delta\varphi]}\right) \qquad (2)$$

Off-axis holography directly extracts $A_sA_r exp^{[-i\Delta\varphi]}$ via standard hologram processing, thus eliminating $A_s^2$ and $A_r^2$.[7] For inline holography we assume operation in the $A_s^2 \ll A_sA_r$ range, a regime that is met during most biological or biomedical applications[14–16], to simplify the discussion. We are thus left with:

$$I_{OA_final} = A_sA_r exp^{[-i\Delta\varphi]} \qquad (3)$$

for off-axis holography and,

$$I_{IL_final} \approx 2A_sA_r cos[\Delta\varphi] \qquad (4)$$

for inline holographies[7]. Irrespective of the approach $A_sA_r$ can be extracted either from the complex field or by strict focus-control and assuming the absence of field curvature to eliminate the impact of $cos[\Delta\varphi]$ from Equation 4. If necessary, $A_r$ can be recovered as the square-root of the measured $A_r^2$ through the median-background approach mentioned above. Ultimately, one is left with a scattering signal, $A_s$, and the challenge is to separate particle-intrinsic scattering properties from optical and measurement-dependent parameters.

A measure of $A_s$ is typically obtained through data processing by, for example, fitting an assumed point-spread-function model and extracting the amplitude, $A_{Fit}$. To relate this amplitude to a particle-specific quantity in the form of a polarizability, $\alpha_{NP}$, it is necessary to correct $A_{Fit}$ for the illumination amplitude at the particle's xyz-position, $A_{illu}(x, y, z)$. Additionally, a potentially spatially dependent $OTF(x, y, z)$ has to be accounted for which might either directly reduce $A_s$ or impact $A_{Fit}$, possibly in a non-linear manner. In simple terms, the extracted quantity, $A_{Fit}$, reports on $\alpha_{NP}$ through a nominally unknown, spatially dependent, normalisation profile, $\nu(x, y, z)$:

$$A_{Fit} = \nu(x, y, z)\alpha_{NP} \qquad (5)$$

Equation 5 suggests a conceptually straight-forward normalisation route. Systematically moving NPs of known size and refractive index and hence polarizability, such as reference materials composed of Au, $SiO_2$ or polystyrene, across the image and then using their position-dependent signal infer the normalisation profile. The difficulty associated with implementing this strategy is the fact that NPs are generally not all of equal size. The mean size of a particle suspension might be known but it is often impossible to know the size of an individual particle under observation. Using spatially dependent scattering signals, from many different single particles, to infer an illumination profile is thus problematic. To demonstrate this problem, we performed a simple simulation of particles distributed on a surface which were illuminated with a Gaussian illumination profile (Supplementary Information 1). Figure 2a compares the amplitudes of simulated holograms for heterogeneous and uniform size-distributions including colour-coded, retrieved signals as obtained through Gaussian fitting of the individual particles. The scattering signals for the former suggest a near-random illumination profile whereas the uniform case suggests a Gaussian-like profile.

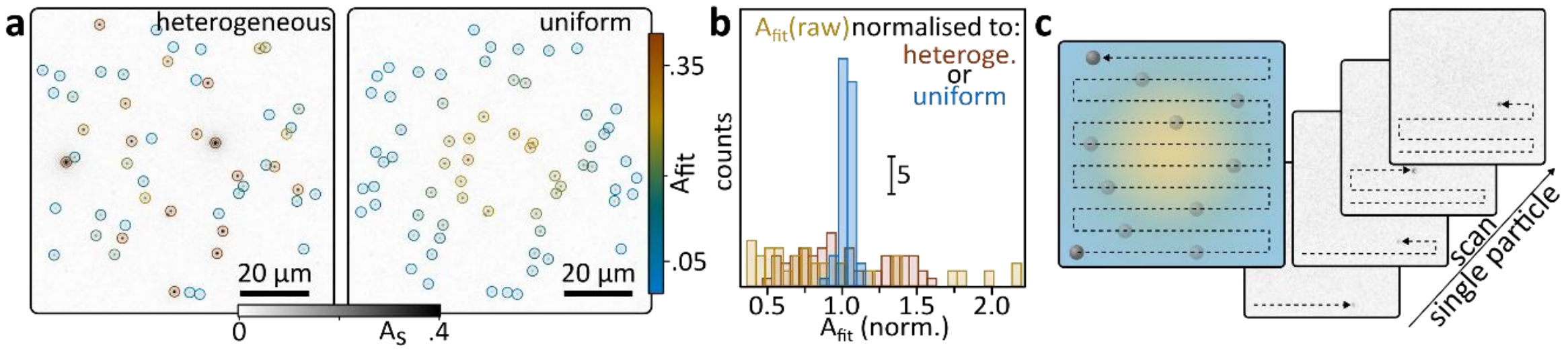

**Figure 2, Scattering signals can be used for self-normalisation, if implemented correctly.** a) Simulated images showing scattering amplitudes for 80 heterogeneous (left) or uniform (right) scatterers immobilised on a 2D surface and scaled with a Gaussian normalisation profile (grayscale). The coloured circles show the recovered particle signal $A_{fit}$. See Supplementary Information 1 for simulation details. b) Histograms of the signals extracted from a) comparing unnormalised, raw, signals (yellow) to those obtained by normalising $A_{fit}$ of the uniform population to a normalisation profile obtained by fitting the scatter plots shown in a) to a Gaussian function. Only the normalisation profile obtained by fitting the uniform case yields satisfactory normalisation results (blue). c) To overcome problems with particle heterogeneity, one could envision systematically scanning a single particle across the entire field of view.

To quantify the problem, we fitted the position dependent retrieved amplitudes to Gaussian profiles thus obtaining normalisation profile estimates using either the heterogeneous or the uniform case. To gauge their quality we then normalised the signals obtained for the uniform case by the two profiles. A satisfactory normalisation profile should recover a near delta-function distribution that is only broadened by shot noise. Figure 2b compares the as-retrieved to the two normalised cases thus highlighting the fact that the normalisation profile obtained for the heterogeneous case is unsuitable for normalisation (see Supplementary Information 1 for a systematic evaluation of this problem).

Real nanoparticle samples are generally heterogenous and the functional form of the normalisation profile is often unknown or ill-defined. As such, experimentally implementing the approach highlighted in Figure 2a,b is difficult. To resolve this problem, one could envision systematically scanning a single particle through the field-of-view (Figure 2c) to sample the normalisation profile while, simultaneously, eliminating the heterogeneity problem. If sampled on a very small grid size, the normalisation profile could be recovered without requiring a functional description. However, systematically scanning a single particle across an image is impractical and cumbersome and the method would implicitly rely on immobilised particles, but many experimental implementations of modern particle sizing target freely diffusing particles to simultaneously measure scattering amplitudes and hydrodynamic diameters[13,17,18]. For these experiments, systematic particle scanning is simply not feasible. Additionally, many particles are typically simultaneously present in a given field of view.

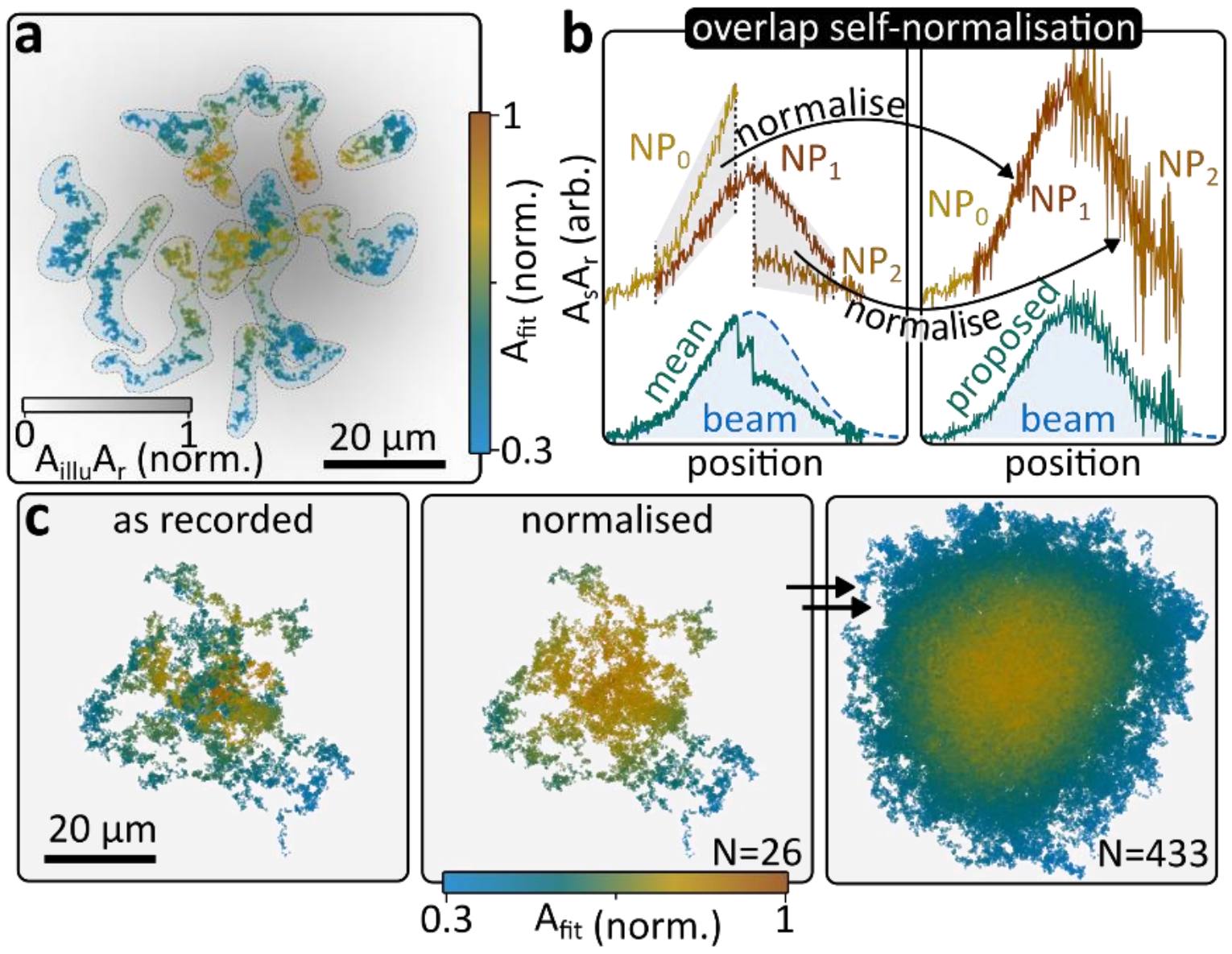

**Figure 3, Implementation of motion-based normalisation profile recovery.** a) The trajectories of 60 nm diameter Au NPs (coloured dots) show particle-specific mean scattering amplitudes but the position-dependent scattering resembles the separately measured beam profile which was acquired analogous to particle scattering, as the hologram-extracted interference term, but with the darkfield stop removed (grayscale image). Individual trajectories are highlighted by area-illustrations (dotted lines). b) Schematic of the proposed approach for scattering cross-section heterogeneity immune self-normalisation based on spatial overlap of single-particle trajectories. Three particles ($NP_0$, $NP_1$, $NP_2$) diffuse through a one-dimensional Gaussian beam profile (dashed line) thus resulting in distinct, particle and position dependent, scattering signals. Following normalisation of the three particle signals using overlapping regions the mean scattering signal changes from a distorted profile with steps (solid green, left) to a profile resembling the simulated beam profile (solid green, right). The shaded grey areas indicate regions used for self-normalising the trajectories with respect to each other. c) 26 randomly chosen trajectories (left) recorded for a suspension containing 40 nm and 60 nm diameter Au NPs can be self-normalised into a continuous scattering profile (centre). Selecting a larger number (N=433) of trajectories recovers the nominal illumination profile as the product of illumination and reference field amplitude (right).

To overcome these limitations we propose an extension of the scanning method which is conceptually outlined in Figure 3. Rather than scanning a particle, we follow the trajectories of many single NPs as they diffuse through the field of observation. Particle-dependent differences in scattering amplitudes are apparent, as indicated by the strongly scattering trajectory overlaying with the weakly scattering trajectory in the centre of the same illumination area (Figure 3a). However, on a trajectory-by-trajectory basis, the single-particle signals report on the independently measured product of illumination and reference wave (Figure 3a, grayscale). In other words, a single trajectory could be used for normalisation, conceptually very similar to manual particle scanning. However, random motion is unlikely to adequately sample the entire field of observation. What is needed is a normalisation approach that uses multiple trajectories of different NPs, to cover the entire area, but is simultaneously insensitive to particle heterogeneity. Toward this goal, we propose a scheme that self-normalises the scattering amplitudes of single particles within their respective 3D spatial trajectories with respect to each other using spatial overlaps as regions where both particles report on identical illumination intensities.

Figure 3b schematically explains the idea, using a simulation of three particles diffusing across a one-dimensional Gaussian beam profile. The particles exhibit different scattering amplitudes which means that a direct signal-mean would fail to recover the beam profile (Figure 3b, left), the experimental heterogeneity problem already observed in Figure 2. To eliminate this issue, we propose to normalise the signals of the individual particles with respect to each other, using mean scattering signals computed in areas of spatial overlap. The rationale is simple: in the absence of heterogeneity, position-dependent signals would be solely a function of $A_{illu}A_r$ and the OTF. If two particles are in close spatial proximity they sample identical $A_{illu}A_r$ and the OTF and a self-normalisation of the particles' entire trajectory based on this overlapping region therefore eliminates size and composition differences. Figure 3b shows that this approach is indeed able to recover the simulated beam profile.

Figure 3c implements the same normalisation in two dimensions, using experimentally measured trajectories for a mixture of 40 nm and 60 nm diameter Au NPs (Methods), particle sizes where gravity-induced sedimentation is quasi-negligible[19]. A set of 26 randomly chosen trajectories shows widely varying scattering signals albeit slightly increased amplitudes towards the centre of the image (Figure 3c, left). We then attempted self-normalisation following a simple strategy (Methods): We first quantised the particle positions onto a 1x1 µm two-dimensional grid and identified the two trajectories with the maximum number of position overlaps. We then normalised these two trajectories using the mean of their respective scattering signals in the overlapping region followed by combining them into a new, longer, trajectory. We then repeated this process until all trajectories were combined into one single spatial network covering the entire area of observation. As can be seen, even 26 trajectories (Figure 3c, centre) already deliver a scattering signal reminiscent of an illumination profile, a marked improvement over the near-random signals (Figure 3c, left). After incorporating the entire dataset, we indeed recovered a profile that closely resembles an expected illumination profile multiplied with a uniform reference wave (Figure 3c, right).

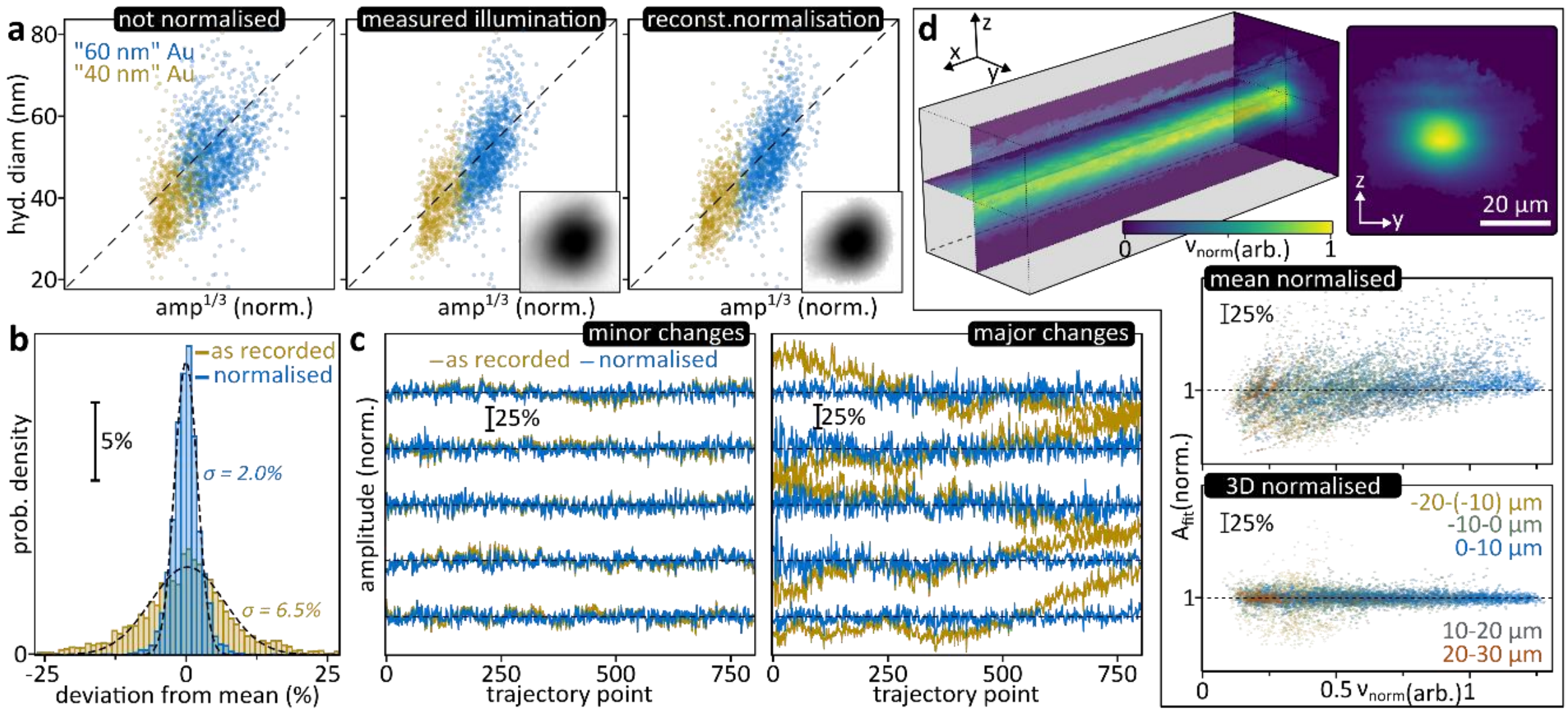

**Figure 4, Experimental benchmarking and application.** a) Hydrodynamic diameter vs cube-root of scattering amplitude plots for individually measured 40 nm (yellow) and 60 nm (blue) diameter Au NPs. Comparisons of results obtained for unnormalised signals (left), those normalised to the directly measured illumination amplitude (centre) or normalised based on a trajectory-based reconstructed normalisation profile (right). The insets show the amplitude of the measured illumination profile (centre) and the reconstructed normalisation profile (right). b) Quantification of signal variations within individual trajectories of diameter 60 nm Au NPs as a measure of normalisation-quality. The histogram shows deviations from the mean value of individual 800-point trajectories. 80 adjacent trajectory points are averaged to eliminate shot-noise contributions thus yielding 10 histogram bins per trajectory. Gaussian fits (black dashed lines) indicate residual amplitude errors of 6.5% (as recorded) and 2.0% (normalised) at one sigma. c) Representative single-particle amplitudes obtained for individual trajectories comparing as recorded and normalisation profile corrected results of trajectories sampling areas of minor variation (left) and drastic changes (right). d) 3D normalisation profile extracted for 60 nm diameter Au NPs using an "invisible" illumination beam based on 90° side-illumination showing the reconstructed xyz-projections (top) as well as trajectory mean normalised scattering amplitudes without (centre) and with (bottom) illumination amplitude normalisation at varying z-distances from the image plane. 8 adjacent trajectory time-points were averaged to reduce shot-noise induced distribution broadening.

At this point, the key question is whether the recovered normalisation profile correctly reflects reality. Answering this question is, at first, non-trivial as the intrinsic particle heterogeneity makes it difficult to provide a ground-truth measurement. Given that we are dealing with freely diffusing particles traditional routes such as comparison to SEM observations of the exact same particles are experimentally out of reach. To nevertheless provide experimental validation we devised multiple approaches that allow comparison with independently measurable quantities that provide different levels of rigorousness: separately measured illumination beamshapes and experiments that explicitly rely on single-particle observations as a function of 3D space, analogous to Figure 2c. The overall results are outlined in Figure 4. Initially, we compared scattering-based recovery with direct measurements of $A_{illu}A_r$, by using a transmission-based darkfield setup where removing the mask allowed a direct measurement of the latter (Methods). This approach yields a somewhat accurate normalisation profile estimate even though OTF- and data-processing-contributions can lead to minor discrepancies[11], an aspect we attribute to the fact that an illumination field measurement only reports on the low-k contributions to the OTF, and the difficulties associated with perfectly positioning a darkfield mask in the Fourier plane.

Figure 4 summarises our validation strategy. Figure 4a compares the hydrodynamic diameters and scattering amplitudes of 40 nm and 60 nm diameter Au nanoparticles (Methods). The unnormalised data show a considerable signal spread, which is reduced to a comparable extent when normalising either to the amplitude of the directly measured beam profile or to the reconstructed normalisation profile. The two profiles are visually similar (Figure 4a, insets), even though they were obtained by distinctly different routes. Because the two are distinct quantities, this agreement is qualitative, but a major discrepancy would indicate a problem with our approach.

For a quantitative test, we evaluated the spatially dependent mean scattering amplitude for single particles based on a trajectory-by-trajectory analysis for 60 nm Au nanoparticles (Figure 4b, Methods). Our underlying assumption is that the particle-intrinsic scattering cross section of a single particle is a constant. Based on this assumption we can use the spatially varying scattering signals of individual particles to evaluate the robustness of our normalisation strategy. To eliminate uncertainties due to particle heterogeneity, we perform a trajectory-by-trajectory analysis where we normalise the scattering signals of each trajectory to their respective trajectory mean. We then computed mean-signals using 80 trajectory-point-bins to eliminate shot-noise contributions. Figure 4b compares the resulting single-particle scattering signals as deviations from the trajectory mean for normalised and as-recorded signals. After normalisation we observe a residual 2% amplitude error (one sigma) on the single-trajectory level, equivalent to a diameter-uncertainty of 0.7%. To assess whether the residual error is due to systematic, position dependent deviations or position-independent noise, such as short-term laser fluctuations and shot-noise, we analysed a few representative single-particle trajectories over a total of 800 time points (Figure 4c). We focused on two extreme scenarios: particles that remained spatially localised in areas with near-constant normalisation profiles and those that explored regions with strongly varying profiles (Figure 4c). The as recorded trajectories show strong variations for the latter cases, with a mean relative standard deviation for the five trajectories of 20.2% compared to the 6.2% for the near-constant case. Importantly, the normalised signals yield similar fluctuations of 7.9% and 5.6%, respectively (all one sigma). The remaining discrepancies between the two cases are likely due to an overall increase in noise towards the regions of low illumination-intensity where particle-localisation is considerably more challenging and signals hence more noisy, aspects that our normalisation routine is not designed to correct for. Importantly, the data presented in Figure 4c comprises trajectories that cover 80% of the total acquisition time of the data underlying the normalisation profile. As such, the absence of systematic normalisation issues suggests that beam pointing instabilities or other alignment issues are negligible.

Thus far, we have focused on normalisations in 2D space, employing an experimental configuration that allowed physically inspecting the illumination field. To demonstrate the applicability of our method, we conclude with an experiment conducted under 90° illumination where the illumination field is undetected (Methods), as outlined in Figure 1b. Figure 4d shows xyz-projections of the reconstructed normalisation profile, which indicates that scattering-based 3D reconstruction might be feasible. To test this hypothesis, we relied on single particle observations, justified by the same assumption that individual particles exhibit constant, position independent, scattering cross sections. Comparing the trajectory-mean normalised scattering amplitudes with and without explicit, position dependent normalisation (Figure 4d) suggests that it is, indeed, possible to reconstruct a 3D normalisation profile based on many individually recorded single-particle trajectories despite the intrinsic particle heterogeneity. More specifically, we observe significant trajectory self-normalised improvements in relative standard deviations comparing normalised (not-normalised) data for the 5 z-ranges: 22.7% vs 32.4% (-20 – -10 µm), 9.5% vs 31.5% (-10 – 0 µm), 6.2% vs 29.3% (0 – 10 µm), 5.1% vs 32.0% (10 – 20 µm) and 4.4% vs 31.0% (20 – 30 µm). The residual error in the -20 to -10 µm z-range, even after normalisation, is likely due to the very low particle counts as well as very low illumination amplitudes that both negatively impacted the reconstruction quality and the overall signal quality.

## SUMMARY AND CONCLUSION

To summarise, we proposed, implemented and validated a methodology that reliably reconstructs a 3D normalisation profile by relying on many single-particle trajectories enabled by stochastic sampling via Brownian motion. At its core, the method eliminates the intrinsic self-normalisation bias, due to particle heterogeneity, through the simple realisation that enforcing identical scattering signals for distinct particles in spatially overlapping regions essentially makes the particles identical. Repeating this self-normalisation procedure over many overlapping regions and many single particle trajectories then covers a large 3D volume that would be, otherwise, impossible to sample with a single particle. Our approach is conceptually related to retrospective flat-field correction methods[20,21], which infer illumination non-uniformity from the image data itself. The key difference is that we normalise overlapping single-particle trajectories against each other, so that each particle's unknown scattering amplitude cancels and particle heterogeneity is removed. Because trajectories are normalised against each other rather than averaged, the approach is insensitive to the detection-threshold bias in weakly illuminated regions, where only strongly scattering particles are detected and an ensemble average would overestimate the local illumination.

The practical benefit of trajectory-based self-normalisation depends on the illumination geometry. If the illumination or reference field reaches the detector, as in inline holography or transmission darkfield with a removable stop, a direct field measurement or median background provides a reasonable first-order normalisation. In these cases our approach is complementary: it offers an independent consistency check and captures contributions that a field measurement misses, namely position-dependent detection and data-processing effects, as well as stray light and back-reflections from other surfaces than the sample plane. These aspects are especially problematic for large field of view observations. Its principal advantage lies in geometries where the illumination never enters the detection path, such as side, oblique or light-sheet-type illumination (Figure 4d). These geometries offer excellent background suppression but have so far lacked a route to quantitative signal normalisation.

We demonstrated our strategy and the problem of particle heterogeneity through serial single- and multiple-particle measurements, initially validating the general applicability through simulation of particles on a 2D surface (Figure 2 and Supplementary Information 1,2) before showcasing how the intrinsic motion of freely diffusing particles enables the same strategy (Figure 3). We then validated our approach through intrinsically more complex scenarios covering illumination profile recovery (Figure 4a), trajectory self-normalisation (Figure 4b,c) and, ultimately, 3D normalisation profile recovery based on experiments which did not allow observing the illumination beam followed by scattering-signal normalisation (Figure 4d). While we explicitly focused on measurements of many individual trajectories, as commonly delivered by our instrument, we believe that sparsely sampled volumes combined with a semi-quantitative normalisation profile model should be equally straight-forward to normalise. As such, our results provide a much-needed signal normalisation framework that is and highly relevant for nanosizing applications and readily translatable to other modalities, such as light-sheet microscopy or inline holography combined with Gouy-phase independent amplitude estimates[22,23], if quantitative signals are of interest. Beyond normalisation, our approach enables accurate and precise spatially-dependent imaging system characterisation that accounts for OTF and signal-processing contributions, based on optical signals that are equivalent to potential analytes of interest. The framework introduced here represents a crucial step towards realising ultrasensitive flow-cytometry based on interferometric imaging rather than darkfield-type point observations.

## METHODS

### Microscope, holographic imaging and data acquisition

Data was acquired using a conventional off-axis real-space holographic microscope[13] (detailed setup sketch provided in Supplementary Information 4) using a 520 nm (PD-01298, *Lasertack GmbH*) laser and a plan fluor objective with a numerical aperture of 0.5 (*RMS20X-PF*, *Nikon*) combination with a 200 mm tube lens and a *Basler dart daA1440-220um* camera as detector with a 100 µs integration time. In all cases, the illumination beam was focused with a 100 mm focal length lens onto the sample. The data underlying Figures 3a,c and Figures 4a-c were acquired at 200 frames-per-second (512x512 px raw image size, 60 videos), the data of Figure 4d at 100 frame-per-second (1024x1024 px raw image size, 340 videos). Each video contained 800 frames with waiting steps equalling the acquisition-duration between videos ensuring that particles moved to considerably different locations prior to acquiring the next video. The probe-to-reference time-delay was adjusted based on a manual translation stage by maximising the amplitude of the interferometric signal. All data pre-processing was conducted as discussed in detail elsewhere[13,17]. Depending on the experiment, the sample was either illuminated in a transmission geometry, with a removable darkfield stop positioned in the centre of the objective's back-focal-plane (Figures 3a,c and Figures 4a-c), or in a side-illumination configuration, under a 90° angle with respect to the collection objective (Figure 4d).

### Sample preparation

#1.5H cover glass was sonicated for 10 min in acetone and 10 min in miliQ followed by drying under a stream of nitrogen and then used for the assembly of sample chambers.

The transmission-chambers consisted of two cover glasses separated by ~400 µm with double-sided sticky tape that were filled with the analyte of interest: 40 nm and 60 nm diameter Au NPs (*BBI Solutions*) diluted to imaging-appropriate concentrations (100-1000 fold dilution of the original OD = 1 suspensions, depending on the experiment).

The side-illumination cells consisted of a single cover glass with two polished fused silica pieces (1x1x8 mm, *Hylasercutting*), separated by approximately 4 mm, acting as in-coupling and out-coupling windows into the NP suspensions contained between the windows. The pieces were attached to the cover glass using an oxygen-plasma activated 50 µm silicone film that had a hole punched into the centre to ensure that the collected signal directly propagated from the liquid environment into the cover glass without passing the silicone layer.

**Data processing**

The interference term is extracted from the as-acquired hologram following standard hologram processing routes analogous to[13]. In brief, the hologram is Fourier transformed, the k-shifted interference term hard-aperture extracted following an inverse Fourier transformation which yields the complex image of interest. Supplementary Information 5 shows a schematic of the workflow. The extracted interference terms were zero padded to 320x320 px (512x512 px videos) or 640x640 px (1024x1024 px videos) and then expanded into a coarse z-stack (14 planes covering ±81 µm in z for Figure 4d, 24 planes covering ±151 µm in z for all other data) using the angular spectrum method followed by particle detection based on a YOLOv8 model trained on a full-physics simulation of the holographic image formation process. Based on the size of the YOLOv8 boxes we inferred a first, coarse, z-position which was then used to obtain near in-focus NPs from the original interference information. Particles with a YOLO confidence score of <0.3 were rejected. The NPs were extracted, cropping an area of 16x16 px followed by multiple rounds of fine-propagation to focus the particle (15 planes each reducing over ±5 µm, ±0.6 µm, ±78 nm and ±10 nm). The numerically obtained z-position yielding the maximum signal-amplitude was used as the true z-position of the particle. The xy-positions were coarsely estimated using an established radial symmetry approach[24]. The particle amplitude was extracted as the maximum of a dramatically up sampled (k-space padding) particle image. Trajectory linking was based on the Python trackpy v0.7 package[25]. The linked trajectories were then used to extract hydrodynamic diameters by first computing the mean squared displacement (MSD) of each trajectory using only the x,y dimensions of the position as they offered favourable localisation accuracy. We fitted first 10 lag times using $MSD(t) = 4Dt + 4\sigma^2$ to extract the diffusion coefficient, $D$, and the localisation error, $\sigma$. The hydrodynamic diameter was then obtained using the Stokes–Einstein relation, $D_h = \frac{k_B T}{3\pi\eta D}$, assuming water at 21 °C ($\eta$ = 0.98 mPa·s).

**YOLO training data generation**

Training data was generated via a simulation of off-axis holograms with parameters matching the respective experiments (numerical aperture: 0.5, wavelength: 520 nm, image size: 512x512 or 1024x1024 px and matching magnification). Each simulated hologram contained 64–128 point scatterers with random lateral positions (offset by at least 20 px from the image edge). The axial positions were random but within the experimental propagation ranges. Amplitude was random (0.5–1) and so was phase (0-2π). The scattered fields were generated in k-space using the angular-spectrum method using a combination of plane-wave phase ramps (x,y) and defocus (z) for positioning the particle within the NA-defined cutoff. The fields were summed and transformed to real space on a zero-padded grid. The resulting scattered field was then interfered with a tilted plane reference wave (amplitude 0.2) to produce the off-axis hologram as a real valued image according to Equation 1. The intensity was subsequently scaled to a 10000 $e^-$ full-well, Poisson noise added, normalised to 255 and then quantised to 8 bits to match the 10000 $e^-$ full-well capacity of the camera at 8 bit image-output. The interference term was extracted according to standard hologram processing (Methods), as for the experimental data. For each hologram, six planes were chosen at random and the field was then refocused at each plane, zero padded to 320x320 (or 640x640 in line with data processing) and saved as an RGB image, with the real and imaginary parts in the green and blue channels. Particles were labelled in a plane if they were located within a suitable z-range with a particle-box centred at their lateral position. The box widths linearly increase with defocus, from 10 px in focus to 40 px at maximum permissible defocus thus encoding the axial position. In 10% of planes, all particles were removed to give particle-free background examples. 5120 holograms were simulated, giving 30720 labelled images. YOLO v8 training was then conducted using 300 epochs, default training parameters and no data augmentation.

**Normalisation profile recovery**

To recover the normalisation profile from the single-particle trajectories discussed in Figure 3 we relied on 2D position-projections to computationally simplify the retrieval process. We quantised the xy/xz/yz-positions using a 1 µm spatial grid and then computed the number of overlapping trajectory points for arbitrarily selected trajectories. If two trajectories overlapped at more than 70 positions, we merged all positions of the two trajectories into a new “trajectory” (other grid sizes and number of overlapping points are also feasible, see Supplementary Information 3). The amplitudes of both initial trajectories were normalised with respect to each other using amplitude means obtained from the overlapping data points. This process was repeated until all trajectories were merged into one trajectory or until a certain number of iterations was exceeded to account for non-overlapping trajectories that were discarded. We repeated the steps outlined above multiple times to ensure consistent reconstruction despite the arbitrary selection of trajectories. The 3D normalisation profile recovery discussed in Figure 4 followed the same, 2D projection-based, strategy which was extended by a 3D interpolation procedure to reconstruct the final volumetric representation and the corresponding normalised scatter plots (Figure 4d).

The 3D reconstruction proceeded as follows. We first smoothed the 2D profiles a Gaussian filter ($\sigma$ = 2 µm), negative values were set to zero, and each projection was normalised to a total of 1. A first estimate of the 3D volume was made by multiplying the three projections together at every voxel and taking the square root. This estimate was then refined in 40 iterations. In each iteration the volume was summed along one axis and compared with the corresponding measured projection. Every voxel was multiplied by the ratio of measured to current projection value. This was done in turn for all three projections. Finally, the volume was smoothed in the dim regions of the beam. For each slice along z, the total intensity was compared with the brightest slice. Slices above 30% of the maximum were left unchanged. Slices below 3% were replaced by a Gaussian-smoothed copy of the volume ($\sigma$ = 3 µm, applied to the logarithm of the volume). Slices in between were blended between the two.

**SUPPORTING INFORMATION**

The Supporting Information contains a simulation of the impact of particle-size heterogeneity on normalisation-profile recovery (Figure S1, Table 1); simulated self-normalisation of heterogeneous particle populations (Table 2); dependence of normalisation-profile recovery on grid size, overlap threshold and acquisition time (Figure S2); schematics of the transmission and side-illumination setups (Figure S3); hologram processing workflow (Figure S4).

## FUNDING SOURCES

D.W. and M.L. acknowledge funding by the European Union (ERC, PIRO, grant number: 101076859). Views and opinions expressed are, however, those of the author(s) only and do not necessarily reflect those of the European Union or the European Research Council. Neither the European Union nor the granting authority can be held responsible for them.

# Supplementary Information:

## Motion-Based Self-Normalisation Enables Precision Nanoparticle Sizing

D-Dré K.J.M.J. Braam[1,2], Daan Wolters[1], Matz Liebel[1,*]

*[1] Department of Physics and Astronomy, Vrije Universiteit Amsterdam, De Boelelaan 1100, Amsterdam, 1081 HZ, The Netherlands*

*[2] Present address: Faculty of Medical Sciences, Radboud University, Geert Grooteplein Noord 21, Nijmegen, 6525 EZ, The Netherlands*

** email: m.liebel@vu.nl*

## Supplementary Information 1: Heterogeneity impacts normalisation,

The intrinsic size-heterogeneity of essentially all nanoparticle populations directly impacts scattering signal based normalisation profile recovery. To highlight this intrinsic normalisation problem we performed systematic simulations.

In brief, we simulated off-axis holograms with parameters matching the experimental configuration used throughout the manuscript (numerical aperture: 0.5, wavelength: 520 nm, 512x512 px, 174 nm/px) but focusing on a pure 2D-surface for sake of simplicity. Each simulated hologram contained a random number of in-focus point scatterers (10-20 range) at random lateral positions but at least offset by 20 px from the image edges. To introduce heterogeneity, we randomly drew particle diameters from a log-normal distribution (mean 60 nm, coefficient of variation 0–80%: 0, 5, 10, 20, 40 and 80%, Figure S1) and then scaled the size to a scattered field amplitude scaling with particle volume ($d^3$) with a "60 nm particle" being assigned a scattering amplitude of 0.3. The amplitude was further scaled by a centred Gaussian illumination profile which corresponds to the normalisation profile in this simulation (full width at half maximum: 300 px). The scattered fields were generated in k-space using plane-wave phase ramps (x,y) within the NA-defined cutoff, summed and transformed to real space, in line with the YOLO simulation (Methods). The resulting field was interfered with a tilted plane reference wave (amplitude 1) to produce the off-axis hologram, which was then scaled to a 3000 $e^-$ full-well and subjected to Poisson noise. The interference term was extracted according to standard hologram processing (Methods).

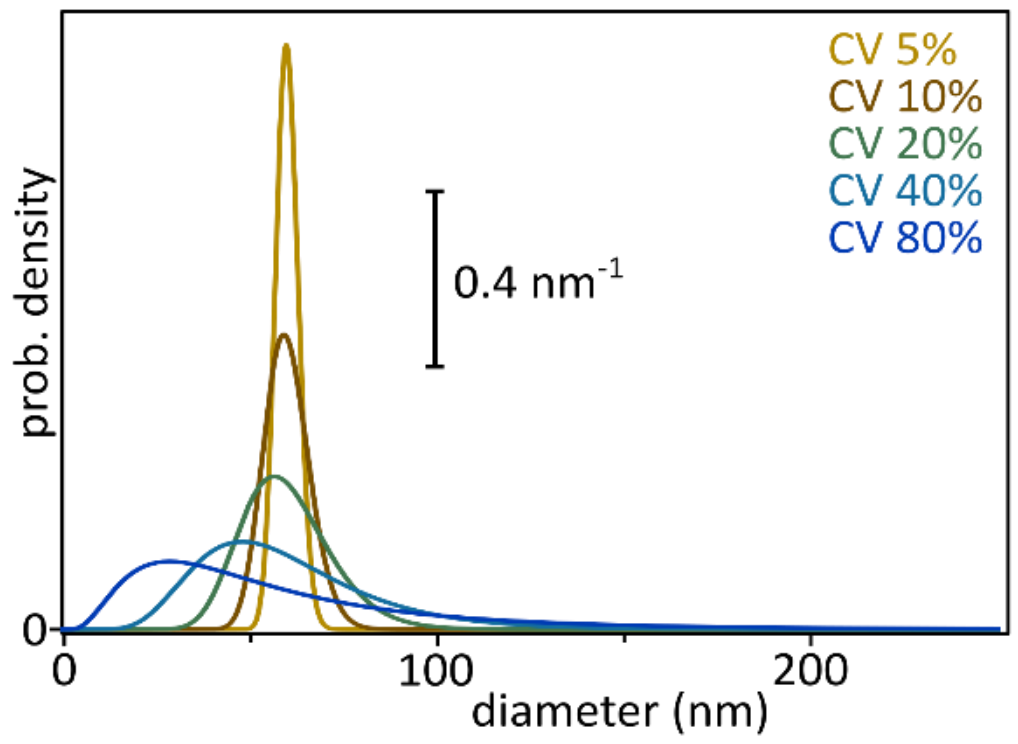

**Figure S1, Particle size distributions used in the simulations.**

Particles were detected as local maxima (15x15 px window) exceeding 8 times the median background amplitude and fitted with a 2D Gaussian (11x11 px). Only fits with a width within ±30% of that of the noiseless point-spread function were accepted to eliminate false positives originating from Airy-ring side lobes of strongly scattering particles. The beam profile was obtained by fitting a 2D Gaussian (amplitude, centre, width) to the fitted particle amplitudes as a function of their lateral position. Multiple simulations containing a total of 20, 100, 200, 400, 800 and 1600 particles were run. Each run was repeated 10 times to allow extracting statistical estimates.

Table 1 summarises the results which shows that the ability to infer a normalisation profile from particle-scattering is highly dependent on the uniformity of the underlying particle population. Observing more particles helps increasing precision but only partially accuracy which is strongly dependent on heterogeneity. Importantly, the simulation and profile-retrieval shown here fits the detection to the known normalisation profile: A Gaussian function. Importantly, experimentally the precise profile shape is generally unknown thus further complicating the retrieval.

| No. NPs | CV 0% | CV 5% | CV 10% | CV 20% | CV 40% | CV 80% |
|---|---|---|---|---|---|---|
| 20 | 298.3 ± 4.4 | 293.7 ± 16.7 | 295.2 ± 22.0 | 318.4 ± 56.9 | 237.4 ± 137.3 | 114.2 ± 125.0 |
| 100 | 298.0 ± 1.4 | 296.6 ± 7.6 | 299.7 ± 8.0 | 310.1 ± 23.3 | 354.8 ± 54.3 | 193.5 ± 163.4 |
| 200 | 298.2 ± 0.8 | 297.9 ± 6.1 | 302.0 ± 11.0 | 311.6 ± 22.5 | 340.0 ± 62.1 | 190.2 ± 200.9 |
| 400 | 298.2 ± 0.7 | 298.6 ± 4.7 | 302.9 ± 9.4 | 306.9 ± 11.3 | 348.9 ± 27.9 | 397.5 ± 120.8 |
| 800 | 298.2 ± 0.6 | 298.3 ± 2.6 | 300.9 ± 5.4 | 309.5 ± 8.3 | 348.3 ± 26.0 | 371.0 ± 104.1 |
| 1600 | 298.2 ± 0.4 | 298.7 ± 1.9 | 300.1 ± 3.6 | 309.0 ± 5.2 | 343.7 ± 20.4 | 349.9 ± 66.2 |

**Table 1**, Estimated full-width at half-maximum of the normalisation profile as a function of number of nanoparticles (NP) observed and coefficient of variation (CV). The correct value is 300 px. Red values indicate systematic deviations, green values are in line with ground-truth. The standard deviations were calculated over 10 runs.

## Supplementary Information 2: Self-normalisation recovers normalisation profiles,

To showcase how self-normalisation allows recovering correct normalisation profiles, we modified the simulation outlined in Supplementary Information 1, focusing on the CV 80% case. We first simulated randomly placed particles until 400 particles were detected and fitted, as above. We then compared this to an idealised "line-scan". Here, we simulated 20 images, each containing a single particle with a diameter randomly drawn from the CV 80% population, placed at 20 equidistant positions along a line through the image centre and scaled with the normalisation profile. One line was simulated every 9°, such that the 20 lines covered the full radial extent of the image, and each line contained one particle exactly in the image centre. Both scenarios used identical detection and fitting. Following fitting, the extracted amplitudes of each line were normalised to the amplitude of the central particle, thus mimicking the self-normalisation approach discussed in Figures 2,3. Lines without a detected central particle were discarded. A Gaussian profile was then fitted to all self-normalised amplitudes and compared to the fit of the randomly placed particles (10 independent trials each). Table 2 shows that self-normalisation recovers the simulated profile, irrespective of the heterogeneity of the particle population, whereas non-normalised populations are unsuitable once heterogeneity starts playing a prominent role. For the CV 80% case the line-scan recovers the simulated profile (FWHM 299.3 ± 2.2 px vs. 300 px simulated), whereas the random population does not (397.5 ± 120.8 px, with 5 of 10 beam fits failing), despite fewer particles being available for the line scan (on average 274 of 400). This supports the statements made based on the experimental observations in the main manuscript.

| Case | Random placement | Normalised lines |
|---|---|---|
| CV 0% | 298.2 ± 0.7 | 298.8 ± 0.4 |
| CV 5% | 298.6 ± 4.7 | 298.6 ± 0.4 |
| CV 10% | 302.9 ± 9.4 | 298.5 ± 0.6 |
| CV 20% | 306.9 ± 11.3 | 298.3 ± 0.6 |
| CV 40% | 348.9 ± 27.9 | 299.2 ± 1.1 |
| CV 80% | 397.5 ± 120.8 (5) | 299.3 ± 2.2 |

**Table 2**, Comparison of recovered full-width at half-maxima for different CV. The correct value is 300 px. Red values indicate systematic deviations, green values are in line with ground-truth. Brackets indicate failed fits. Red values indicate systematic deviations, green values are in line with ground-truth. The standard deviations were calculated over 10 runs.

## Supplementary Information 3: Impact of recovery parameters,

Throughout the manuscript we employed a trajectory overlap of 70 points and a grid size of 1000 nm but other parameters are equally applicable as summarised in Figure S2. Independent of the grid size and number of overlapping trajectory points we observe near-identical normalisation profiles. Repeatedly running the algorithm over a given dataset yields consistent normalisation profiles (Figure S2a) with mean normalised standard deviations, computed based on 10 runs, being <2% (one sigma).

The cross-cuts over the centre of the image reveal grid size (Figure S2b) and point-overlap (Figure S2c) independent recovery. Splitting the data in two subsets reporting on the first and last half of the videos recorded, further, reveals near identical normalisation profile recovery (Figure S2d) with the differences at the edges of the profile being due to the lack of trajectories being present in the respective low-amplitude regions.

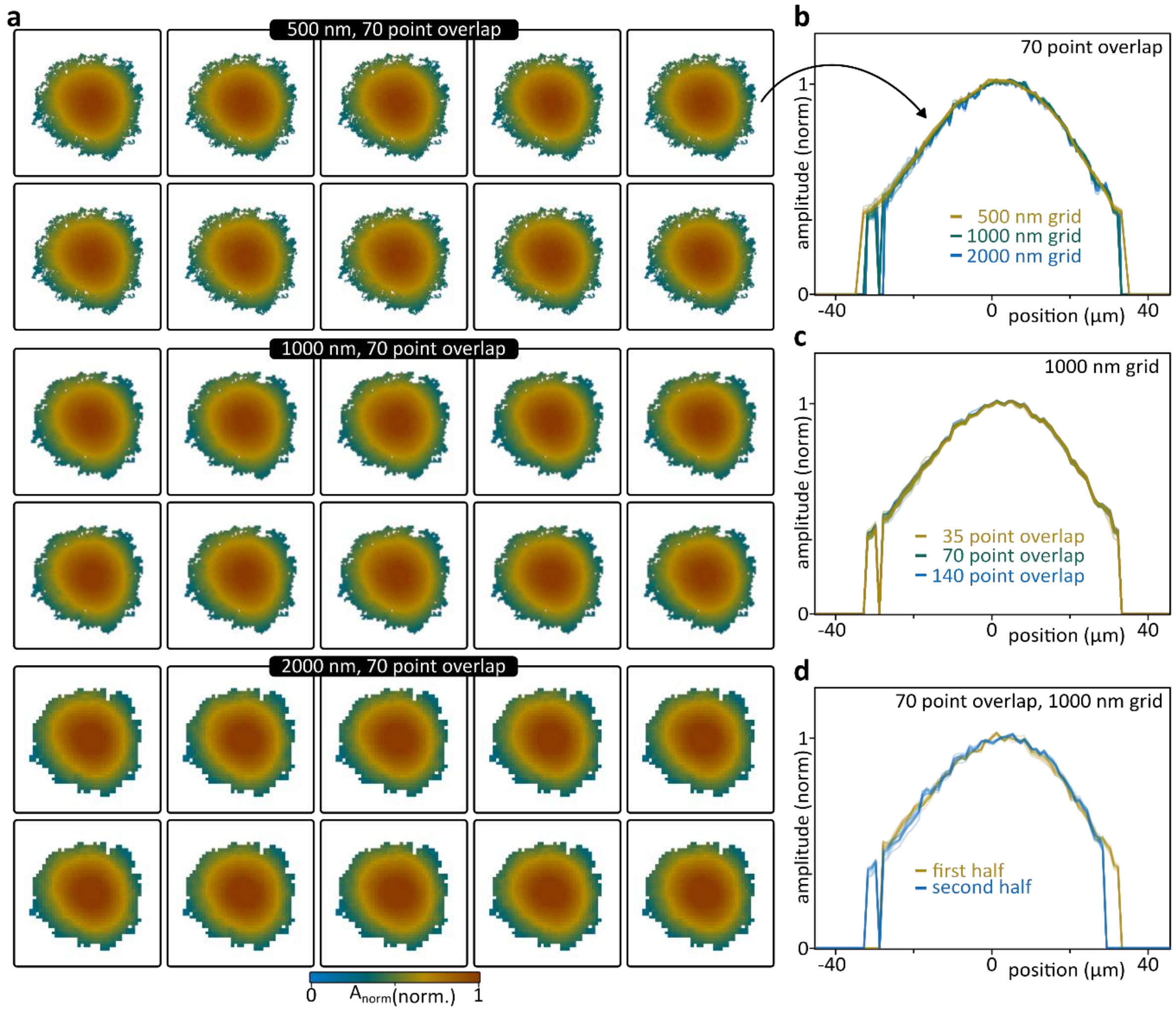

**Figure S2, Recovery parameters only marginally impact normalisation profile recovery.** a) Comparison of normalisation profiles recovered for the same data set and recovery parameters for 10 consecutive runs of the retrieval algorithm. b) Cross-cut comparison across the centre of the images shown in a) shows near-identical normalisation profiles irrespective of the grid size. c) As b) but for varying number of overlapping points. d) Cut identical to the above but for normalisation profiles recovered for the first (last) 30 videos recorded. The data in the Figure is identical to the dataset used for Figure 3a,c. All profile shown were recovered using 433 trajectories of 800 points each. The comparison using half the data used 226 (first half) and 207 (second half) trajectories for their respective beam profiles.

## Supplementary Information 4: Schematics of setups used,

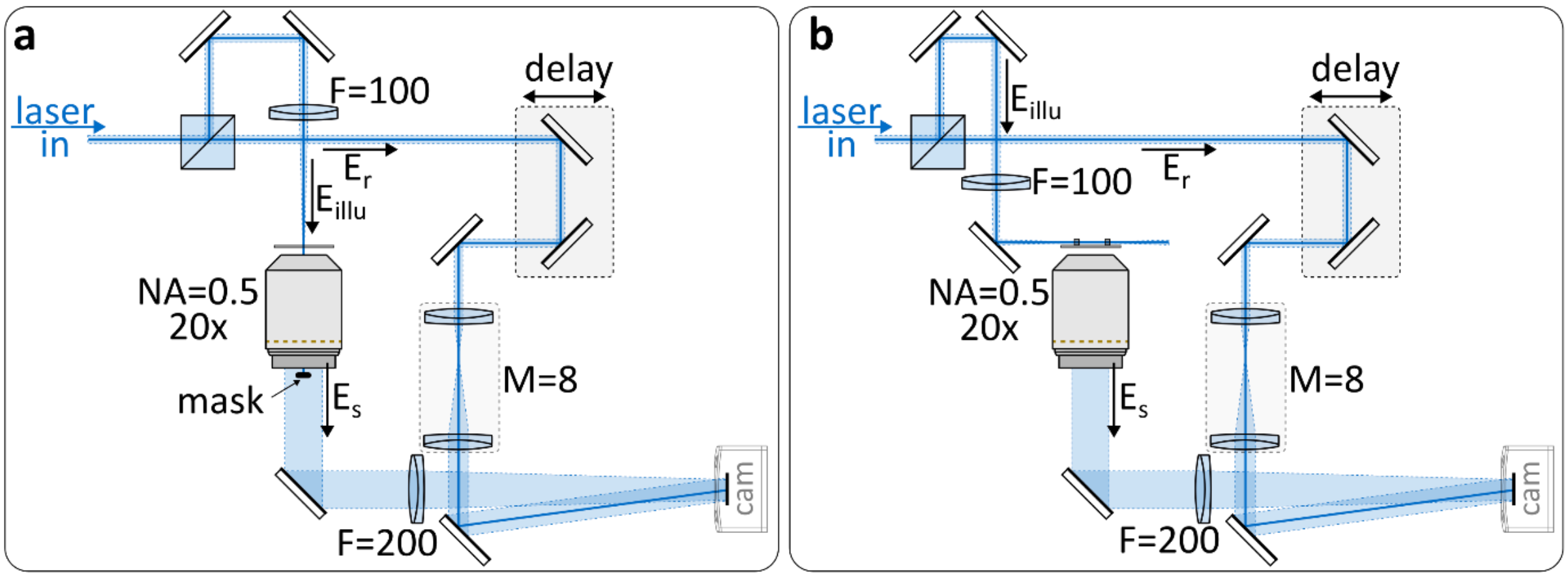

**Figure S3**, Setup schematics. a) Transmission-based "darkfield holography" implementation where the illumination beam passes through the objective and is then blocked by a darkfield mask place in the vicinity of the microscope objective's back focal plane. b) Side-illumination setup where the illumination beam interacts with the sample but does not pass through the objective.

Figure S3 summarises the two experimental configurations used based on either transmission- (Figure S3a) or side-illumination (Figure S3b) with the former using a darkfield mask to prevent the illumination light from reaching the camera. As discussed in the main manuscript, a 520 nm (PD-01298, *Lasertack GmbH*) laser was used in combination with a plan fluor objective with a numerical aperture of 0.5 (*RMS20X-PF*, *Nikon*) and a 200 mm tube lens. a *Basler dart daA1440-220um* camera serves as detector with a 100 µs integration time. The illumination beam was focused with a 100 mm focal length lens onto the sample. A manual translation stage adjusts the signal to reference time delay to ensure interference. Prior to reaching he camera, the reference beam is expanded eight-fold by means of a telescope. The experiments presented in Figure 3a,c and Figure 4a-c of the main manuscript were recorded using a setup analogous to Figure S3a, the data in Figure 4d was acquired with a setup analogous to Figure S3b.

## Supplementary Information 5: Hologram extraction,

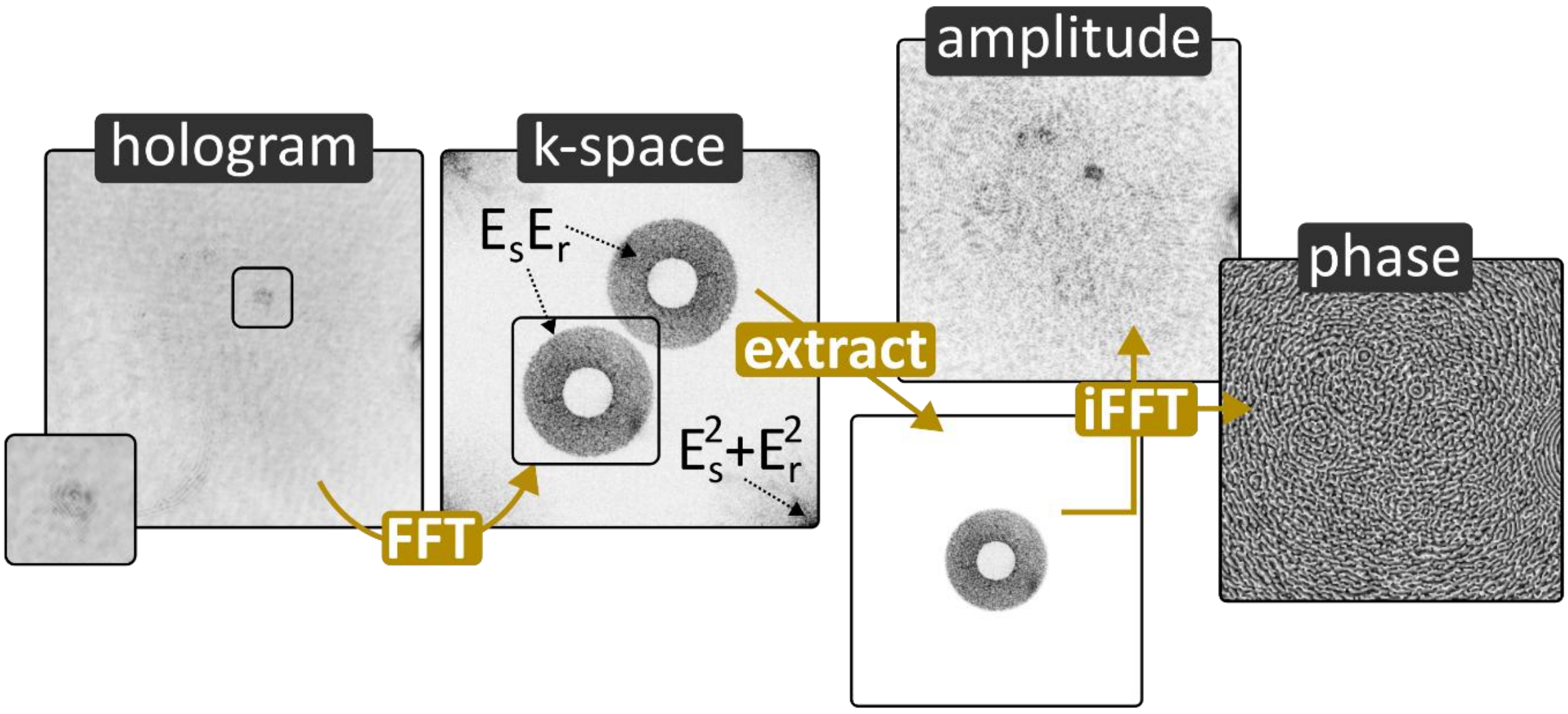

**Figure S4, Hologram extraction workflow.** The k-space representations have k=(0,0) at the corner. The extracted interference term requires an iFFT-shift prior to the inverse Fourier transformation (not shown). The zero-padding of the extracted k-space is for representation purposes only.

Figure S4 summarises the hologram extraction workflow based on data used in the context of Figure 3a,c. In brief, the as-recorded hologram is Fourier transformed (FFT) to separate the interference terms from the amplitude square terms. One inference term is hard-aperture selected by cropping it from the k-space followed by multiplication with a circular mask whose size is determined by the image magnification. The extracted k-space is Fourier shifted, to centre it at k = (0,0) followed by an inverse Fourier transformation (iFFT) to yield the complex field of interest which can then be used for further data processing. Based on these fields it is then possible to propagate to cover the large, sparsely populated, volume of interest followed by particle detection and trajectory linking over many images, e.g. time-lapse recordings.